\documentclass[review]{elsarticle}

\usepackage{lineno,hyperref,enumitem}
\modulolinenumbers[5]

\journal{Nuclear Instruments and Methods in Physics Research Section A}

\begin{document}

\begin{frontmatter}

\title{GGS: a Generic Geant4 Simulation package for small- and medium-sized particle detection experiments.}

\author{Nicola Mori}
\address{INFN sezione di Firenze, via G. Sansone 1, I-50019 Sesto Fiorentino, Florence, Italy}
\ead{mori@fi.infn.it}

%
%
%

\begin{abstract}
The Generic Geant4 Simulation (GGS) is a package designed to speed-up the realization and deployment of Monte Carlo simulation software 
based on Geant4, for small- and medium-sized high-energy experiments. For many common use cases, the task of setting up a full-featured 
simulation of the detector is reduced to the definition of the detector geometry, by providing generic and reusable implementations of 
the mandatory Geant4 user classes (particle generation, scoring, output etc.). Extensibility is provided by a simple plugin system that 
allows replacing of the generic implementations distributed with GGS with custom ones. These features make it especially suitable for 
cases where limited manpower, like during preliminary detector design studies, can severely limit the scope of an R\&D program.
\end{abstract}


\end{frontmatter}


\section{Introduction}
Monte Carlo simulations of radiation-matter interactions play a crucial role across the whole spectrum of experimental high-energy physics 
activities. Given the variety of the involved physical processes and the intrinsic difficulty in modeling them (due to e.g. the necessary 
mathematical and computational approximations) the implementation of a full set of such processes within a computer code and its validation
require a huge effort. Thus it is practically impossible to develop a simulation code from scratch for every experiment. The community has 
tackled this issue by developing generic and reusable software packages that provide an implementation of the physics (and of related 
needed facilities like random number generators) to be used with different detector geometries. These packages constitute a solid common 
ground upon which the different collaborations can build up their detector simulation, saving development time and relying on tested and 
validated implementations of critical components.

Currently, the most sophisticated and popular among such software frameworks in the high-energy community are Geant4 \cite{Allison2016} 
and FLUKA \cite{Battistoni2015}; thanks to their flexibility, they have been adopted also in other fields, from medical to nuclear physics 
and space applications (see e.g. \cite{Faddegon2020,Botta2013,Lattuada2017,Santin2006,Lee2007} for a few examples of applications to these 
fields). Such wide usage has been achieved, among other things, by leaving the implementation of experiment-specific components like the 
detector geometry, the content and format of the output files etc., to the users, while providing the infrastructure for using these 
specific components within the framework (for example, Geant4 is written in C++ and makes use of polymorphism to this end). This approach 
allows for extreme customizability and optimization and thus for a great range of possible applications, but has also noticeable drawbacks, 
the most evident being the overhead imposed by the implementation of such components. Typically, these implementations rely either on a 
declarative DSL (Domain Specific Language) or on an API (Application Programming Interface) for a general-purpose programming language like 
C++. In both cases, the user must first learn the DSL or the API and then implement the needed components: the first part can be very 
time-consuming and sometimes technically challenging but is usually related only to the first approach to a toolkit, while the second one is 
generally unavoidable when implementing a new simulation code. This overhead may be significant and thus drain a substantial amount of 
resources that in some cases (e.g. a small collaboration with limited manpower, or an individual wanting to perform some preliminary studies 
about a new detector concept or design) can drastically limit the scope of the simulation.

Another issue related to customizability is fragmentation. The freedom to implement the user components often leads to different solutions 
produced by different developers (and also by the same developer), which prevents the formation of a common ground for sharing knowledge 
and reusing code. As a trivial example, a person working in two different collaborations will likely have to learn how two simulations are 
run, how two different data output formats have to be read and analyzed, and so on. Usually none of the two custom tuned solutions can be 
easily adapted to a third use-case, which would then need a new design and implementation.

Different solutions have been developed to mitigate these problems. The GAMOS framework \cite{Arce2014} is focused on giving the user the 
possibility to run a Geant4 simulation without the need to write C++ code. It uses a declarative, proprietary DSL to define the 
detector geometry, and provides many reusable facilities like particle generators mainly catering to medical physics; a plugin system 
allows to extend its functionalities to other energy ranges and use-cases. In the high-energy field, the LHC collaborations have produced 
many advanced simulation frameworks (see e.g. \cite{Aad2010,Clemencic2011}), but they are not generally available nor designed to be 
reusable. A recent effort \cite{Siddi2019} in producing a generic and reusable framework shows a promising potential. For space 
applications, there are available simulation packages; however, these are usually aimed at specific use-cases like shielding \cite{Lei2002}, 
space environment studies \cite{Santin2005} or well-defined energies/particles \cite{Zoglauer2006}.

The Generic Geant4 Simulation (GGS) software is an experiment-independent package that has been designed to ease and speed-up the 
realization and deployment of Monte Carlo simulations for particle-detection experiments based on Geant4. It has been originally conceived 
in the field of direct cosmic-ray detection and is aimed at small and medium-sized detectors and collaborations. It is based on a generic 
and reusable implementation of all the Geant4 user classes (except for detector geometry) that may be used in many typical use-cases, 
reducing the required user implementation to the bare minimum (i.e. the detector geometry itself). It features an output format based on 
Root files \cite{Antcheva2009} that is automatically adapted to any possible detector configuration, plus a set of readout routines for 
offline analysis. A simple plugin system allows to customize almost all of the different simulation aspects, from particle generation to 
output, whenever the default implementations are not suitable. Application steering is performed using the Geant4 messenger system. 

In the following, section \ref{sect:requirements} gives an overview of the design requirements of GGS. Section \ref{sect:architecture} 
details some of the most significant architectural choices and implementation details. In section \ref{sect:resources} some references about 
available online resources for GGS are given. Finally, some real-world use cases are briefly described in section 
\ref{sect:usecases}.

\section{Design requirements}
\label{sect:requirements}
The main idea behind GGS is to offer a robust starting point for all those situations where a quick deployment of a Geant4 Monte Carlo 
simulation of any kind of detector is needed. In these cases, which include e.g. a detector prototype response to a beam of test particles 
or the preliminary performance studies of different possible instrument configurations, usually a limited accuracy will suffice. Thus it 
may be convenient to rely on available general simulation components (e.g. particle generation, physics processes, data output etc.) that 
can provide a reasonably accurate first set of results while reducing the necessary user development to a minimum, also making it possible 
to profit from previous experiences with such components from different use cases. This last aspect is further enhanced when the output 
format is kept the same (as much as possible) for different detector geometries, so that the different offline analyses can be implemented 
using the same data readout procedures. These ideas then translate into the following requirements:
\begin{enumerate}
\item it must be possible to simulate any detector geometry;
\item the detector geometry, being experiment-specific, must be specifiable by the user externally (in other words, no reference to any 
particular geometry must be present in the GGS code base);
\item there must be generic ``standard'' implementations of all the other mandatory (particle generation, physics list) and optional (user 
actions, scoring) Geant4 user classes which are suitable for a vast class of simulation cases;
\item the output format must be the same for any simulated geometry.
\end{enumerate}
A software solution meeting these requirements can reasonably cope with the preliminary phases of simulation studies in many cases; 
however, it falls short as soon as the need for more detailed results arise. In these cases it must be possible to replace the standard 
implementations with custom, experiment-specific ones, or to add new functionalities besides the standard ones. Being experiment-specific 
these custom components must be externally developed and then plugged into the standard GGS simulation, to keep the GGS code base 
experiment-independent. These considerations lead to the following additional requirements:
\begin{enumerate}[resume]
\item it must be possible to replace any standard component with custom ones, and also to use custom components together with standard ones
where it makes sense;
\item as for the geometry, all the custom components must be developed externally;
\item custom components must be usable without needing any modification to the GGS code base;
\item the format of the output file must be modular, so that any additional information generated by custom components can be saved 
together with the standard one without breaking the format.
\end{enumerate}
This enlarged set of requirements ensures that after the first batch of simplified studies is performed there is the possibility to add 
more details to the simulation without breaking any of the already existing code. They also extend the application of GGS to those cases 
that are not covered by the default components, although at the price of requiring custom implementations from the very beginning; using GGS 
can however be worthwhile also in these cases, should any of the standard components still be viable.

Given the focus on small experiments, no specific requirements about multithreading are made. The Geant4 multithreading model 
\cite{Allison2016} parallelizes the simulation of different events using threads, which share a single copy of the most memory-consuming 
entities (the geometry and the tables used by electromagnetic physics processes). This model makes an efficient use of computing cores in 
memory-constrained scenarios, which is the typical use-case for e.g. LHC experiments. For smaller detectors the memory used by the geometry 
is usually not an issue on modern machines, so a parallel computing model based on multiprocessing (i.e. many instances of a single-threaded 
program) has practically no disadvantage with respect to one based on multithreading (i.e. a single instance of a multi-threaded program), 
with the advantage of a much simpler implementation of the simulation program.

Since knowledge of C++ has become a very common skill in the communities at which GGS is targeted (mainly, cosmic rays and particle 
physics), being routinely taught also in undergraduate courses, there is no requirement about making GGS usable by people without its 
knowledge, unlike similar packages (e.g.\cite{Arce2014}) that are explicitly designed for this purpose. In addition to this, in the most 
favorable case, the user only has to specify the detector geometry in order to be able to run a full simulation, and thus has to be familiar 
only with this part of the Geant4 C++ API. For these reasons, the knowledge of C++ and of the Geant4 API are not considered general blockers 
to the use of GGS; for those cases in which this can become a stumbling block it is possible to define the geometry with a GDML file 
\cite{GDMLweb}, as described in section \ref{sect:geometry}, making it possible to set up a full GGS simulation with standard components 
without the need to write any C++ code.

\section{Architecture}
\label{sect:architecture}
\begin{figure}[htbp]
\begin{center}
\includegraphics[width=12cm]{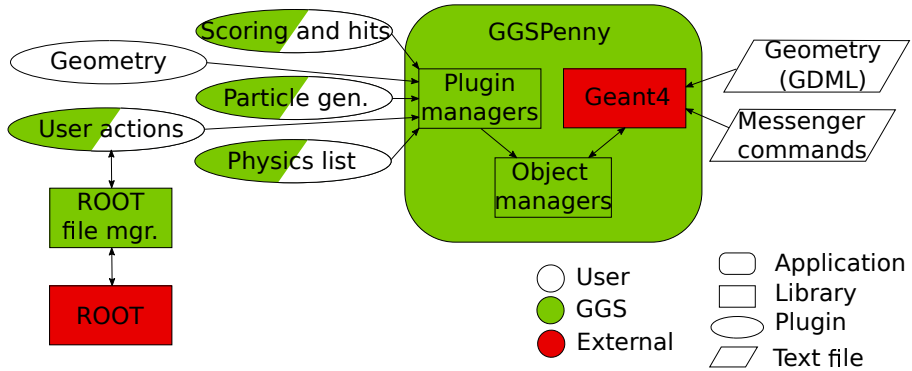}
\end{center}
\vspace{-0.5cm}
\caption{Scheme of the GGS architecture.}
\label{fig:architecture}
\end{figure}

GGS consists of a set of C++ classes inheriting from Geant4 base classes, which constitute the standard implementation of user classes, an 
executable application, a set of facilities like e.g. a simple plugin infrastructure, and data readout classes for offline analysis. An 
overview of the GGS architecture is shown in figure \ref{fig:architecture}.

In the following, the different components are described in details.

\subsection{Plugins and configuration}
\label{sect:plugins}
A key requirement for GGS is to be able to use custom components developed as C++ classes by users in downstream projects. This is 
accomplished by means of a simple plugin system based on global proxy objects and registration macros, and of a set of object manager 
classes that act as object factories.

When developing a custom class, the developer must place a call to a registration macro provided by GGS in the same source file; this 
macro defines a builder function that creates and returns an object of the custom class, a thin proxy class that in its constructor 
registers this builder function into the object manager, and a global instance of this proxy class. When the custom code is built into a 
shared library and then dynamically loaded at runtime by the GGS plugin system, the global proxy object is built and thus the custom class 
builder is automatically registered into the object manager. Each object manager holds a (class name, class builder) map for all the 
registered classes, so it can build an instance of a custom class by means of a string with the name of the class to be built. The object is 
returned by the builder function as a pointer to its base class, and can then be manipulated polymorphically without needing to know its 
exact type; in particular, this means that the headers for the custom classes are not needed, so these classes can be developed in different 
projects.

Any number and kind of custom classes (with some exceptions, see the next sections) can be placed in the shared library, since at runtime 
it will be possible to specify which ones have to be instantiated and used, by feeding the object managers with the relevant class names. 
To this end, the Geant4 messenger system is used to define which objects should be built and used, and more generally to configure the 
simulation runs. It is based on a declarative DSL that can be extended programmatically to add new commands, and provides all the usually 
needed facilities like the string-number conversions and the handling of the units of measurement for specifying physical quantities. 
Commands are grouped in folders, and the GGS ones are all placed under the \verb|/GGS| root. GGS commands can be used to create and 
configure at runtime the instances of the user classes contained in plugin libraries. In fact, also most of the GGS standard classes are 
handled with this mechanism, so that standard and custom components are treated uniformly (the only difference being that standard component 
libraries are loaded automatically while those for custom components have to be specified when launching the simulation application).

\subsection{Geometry}
\label{sect:geometry}
The definition of the detector geometry is referred to the user. GGS allows for using both the Geant4 C++ geometry definition API and GDML 
files \cite{GDMLweb} which are natively supported by Geant4. Regarding the C++ API, there is no fundamental difference in defining a 
detector geometry for GGS or for another Geant4 application, with two exceptions. The first one is that the geometry must be registered 
using a macro, as described in section \ref{sect:plugins} (the geometry plugins are handled in a special way, so that in each library only 
one geometry class must be present at most). The second one is that sensitive detectors must not be associated to the logical volumes in 
the geometry construction code, since this task is performed by GGS at runtime, ensuring an easy on/off switching of sensitive volumes and 
changing the type of sensitive detector by means of messenger commands. Due to this feature, geometry classes used in GGS must inherit 
from {\it GGSVGeometryConstruction} and not directly from {\it G4VUserDetectorConstruction}.

Inheriting from {\it GGSVGeometryConstruction} also offers the possibility to define an instrument acceptance. The user can define 
a geometry-specific routine for determining whether a generated primary particle is inside the acceptance or not, and if not GGS will 
immediately discard it and generate another one. In this way the computation is optimized by effectively simulating only the particles 
generated inside the detector acceptance.

Another feature of {\it GGSVGeometryConstruction} is the possibility to build the geometry according to some parameters defined in a 
geometry configuration file provided by the user. In this way it is possible to customize the geometry at runtime, which can be useful to 
e.g. remove a sub-detector or change its dimensions. For maximum flexibility GGS leaves the handling of geometry parameters entirely to the 
user.

Finally, {\it GGSVGeometryConstruction} allows to export the values of the geometry parameters to the simulation output files for use 
during the offline analysis. Again, complete freedom is left to the user about defining which parameters to export: all or just some 
of the ones given by the geometry configuration file, some combinations of them etc. This is a very limited way to make the simulation 
geometry available during offline analysis, especially if compared with more powerful and dedicated tools \cite{Petric2017}, but for 
simple cases it constitutes a handy solution. A more complete alternative is mentioned in section \ref{sect:useractions}.

\subsection{Particle generation}
GGS provides the {\it GGSGeneratorAction} class from which particle generator actions can inherit. This class adds a couple of features to 
the standard Geant4 {\it G4VUserPrimaryGeneratorAction} class: one is the possibility for generator actions to export the generation 
parameters (similarly to the detector geometry feature described in \ref{sect:geometry}), the other is the number of events that have been 
generated and then discarded because they failed the instrument acceptance cut as defined by the detector geometry. GGS provides four kind 
of generator actions:
\begin{itemize}
\item a single-particle generator action: it makes use of the {\it G4Gun} generator and implements some features that were 
originally conceived for direct cosmic-ray detection experiments but that can be useful in many other scenarios. The most prominent are: 
random generation point sampled from volumes (parallelepiped) or surfaces (sphere cap, plane), random generation direction, random 
energy spectrum (flat or power law), generation inside the instrument acceptance (see \ref{sect:geometry});
\item a HEPEVT generator action for particles specified in an input text file in the standard HEPEVT format;
\item a generator action extending the functionality of the HEPEVT generator by allowing to define in the input file also the generation 
point and time for each primary particle;
\item a generator action wrapping the powerful Geant4 General Particle Source (GPS) generator that allows to customize almost every aspect 
of particle generation using the Geant4 messenger system.
\end{itemize}
Custom particle generators are supported by the usual plugin mechanism.

\subsection{Physics list}
The creation and tuning of a physics list is a lengthy and involved process that is typically tailored on experiment-specific needs. For 
this reason the provision of physics lists is considered to be outside the scope of GGS. Instead, GGS allows to use either one of the 
predefined Geant4 lists like FTFP\_BERT, QGSP\_BERT etc., or a custom list created by the user. In the former case, the physical accuracy 
of the results obtained with GGS is that of any other simulation software making use of the standard Geant4 physics lists. There is a 
vast amount of available validation studies for the physical process implemented in Geant4 (see e.g. \cite{g4pub} for a comprehensive list, 
and \cite{g4val} for a detailed and up-to date comparison of different Geant4 versions) that can be used as a reference for the accuracy of 
GGS with standard Geant4 physics. User-defined physics lists rely on the GGS plugin mechanism, and as for the geometry plugins also the 
physics list plugins must contain only one physics list class.

\subsection{Hits and scoring}
\label{sect:hitsandscoring}
The scoring mechanism implemented in GGS is based on the plugin and configuration infrastructures described in section \ref{sect:plugins}. 
Sensitive detector (SD) code is placed in plugin libraries, and SD objects are built and associated to logical volumes at runtime by means 
of messenger commands. This simplify the activation of a volume and changing the SD. Multiple SDs can be assigned to the same volume, 
if needed, to produce different kind of hits for the same volume.

The user can define custom SDs and hits using the standard Geant4 API for maximum flexibility. For typical use cases, a generic 
ready-to-use SD is available as the {\it GGSIntHitSD} class. For each sensitive volume it creates hit objects with a varying degree of 
detail:
\begin{itemize}
\item position hits: a position hit records the energy deposited by a single particle during a single simulation step inside a sensitive 
volume. This is the most fine-grained hit type;
\item particle hits: a particle hit records the total energy deposited by a single particle while traversing a whole sensitive volume, 
together with some ancillary information as the particle PDG code, the entrance and exit points, the entrance energy and so on. The particle 
hit can be thought of as the sum of all the position hits for that particle in the given volume;
\item integrated hits: an integrated hit records the total energy deposited by all the particles traversing a sensitive volume. It can be 
seen as the sum of all the particle hits for that volume.
\end{itemize}
The hits are stored in memory hierarchically: each integrated hit contains a list of its corresponding particle hits, and each particle hit 
contains a list of its position hits. To balance the level of detail and the memory occupancy, the type of hits to be used can be defined on 
a per-volume basis at runtime with messenger commands. 

Most of the basic scoring needs are reasonably covered by the set of hits described above. The default implementation of any hit type 
can be replaced with a custom one separately for each sensitive volume if additional scoring information (e.g. the absorption of transition 
radiation X-ray photons) other than the deposited energy is needed.

\subsection{User actions}
\label{sect:useractions}
The GGS infrastructure for user actions is based on plugins and on configuration via messenger commands like some of the other components 
seen so far. Apart from the possibility for the user to implement the user actions using the Geant4 API, GGS provides an additional API 
layer with the {\it GGSUserAction} class. By inheriting from all of the Geant4 user action classes ({\it G4UserStackingAction }, {\it  
G4UserTrackingAction}, {\it  G4UserSteppingAction}, {\it  G4UserEventAction} and {\it  G4UserRunAction}), a class that inherits from it 
can group in one single action all the functionally correlated routines that have to be executed at different stages of the simulation. 
This makes it easier to propagate information to different simulation stages and to add a given functionality to the simulation run by 
adding a single user action with a messenger command. Multiple user actions can be created at once with messenger commands in a simulation 
run, making it possible to customize at runtime the output data set by choosing the set of the desired actions among the available ones.

GGS comes with some ready-to-use user actions. Many of them output information to a Root file; those of most common usage save on file 
the hit objects produced by the {\it GGSIntHitSD}, the Monte Carlo truth about the primary particles for each event, the information about 
the hadronic interactions of the primary hadrons, and the full simulation geometry in TGeo format by leveraging the Virtual Geometry Model 
(VGM) library \cite{Hrivnacova2008}. Different and independent user actions can share the same Root output file and the same Root event 
tree for saving event data by means of a centralized Root file management facility.

\subsection{The GGSPenny executable}
\label{sect:GGSPenny}
The GGS application which runs the Monte Carlo simulations is called {\it GGSPenny}. It is a standard Geant4 application implementing the 
usual flow for configuring, initializing, running and finalizing the simulation using the GGS components described in the previous 
sections. It provides some command line flags to specify the plugin libraries to be loaded, the physics list to be used, the random 
seeds to initialize the random engine, and so on.

{\it GGSPenny} can be run either in batch mode or in interactive mode, the latter being useful to visualize the detector geometry and 
manually enter the messenger commands. The visualization relies on the standard Geant4 visualization infrastructure and can be configured 
and steered using the standard visualization commands.

\subsection{Offline data readout}
A set of data readout classes to ease the readout of the simulation output from Root files during offline processing is available with the 
GGS code base. Each readout class reads the information saved by one user action during the simulation run; this design allows to easily 
extend the readout infrastructure when a new user action is created, by implementing its corresponding reader class. The class {\it 
GGSTRootReader} takes care of instantiating the reader classes when requested and to keep all of them aligned at the same event of the 
shared event tree.

\subsection{Utilities}
\label{sect:utilities}
Besides the main components described in the previous sections, GGS offers some utility applications for common needs encountered when 
dealing with Monte Carlo simulations of particle detectors:

\begin{itemize}
\item geometry conversion: the {\it GGSWolowitz} executable converts the plugin geometry into the GDML or TGeo formats. For the former the 
internal Geant4 converter to GDML is used, while for the latter the VGM library \cite{Hrivnacova2008} needs to be present in the system. 
This allows to easily interface the GGS geometry with visualization tools using these popular geometry description formats;
\item event display: the {\it GGSLeonard} executable is a generic 3D event viewer based on the Event Visualization Environment (EVE) 
\cite{Tadel2010}, capable of displaying the detector geometry and the energy deposits in the different sensitive volumes as read from a 
Root output file. It has limited functionalities but provides a handy and ready-to-use solution for a task that usually requires a 
considerable amount of code development.
\end{itemize}

\section{Resources}
\label{sect:resources}
The GGS source code is freely available under the GPLv3 license. The code is maintained in a git repository managed by GitLab 
\cite{GGSrepo} and hosted by the Italian Istituto Nazionale di Fisica Nucleare (INFN). The Gitlab web page provides a centralized access 
point to the code and to all the resources needed by users and developers, like the user's guide and the Doxygen documentation. GitLab is 
used also to prepare and host tagged code releases, to track bugs and development, and to test the code builds on different flavors of Linux 
and macOS by means of the continuous integration service.

\section{Use cases}
\label{sect:usecases}
A number of small projects and some small- and medium-sized collaborations have used and are using GGS for their Geant4 Monte Carlo 
simulation codes.

Preliminary studies about innovative detector designs or measurement principles have been conducted by individuals and small research 
groups of a few persons without needing a deep knowledge of the Geant4 API thanks to the usage of the default components provided by GGS 
(see e.g. \cite{Duranti2021}), profiting also of the adaptive, robust and ready-to-use Root output and offline readout routines. It is also 
remarkable that even in the context of bachelor degree theses the usage of GGS has allowed individual undergraduate students to build up a 
Geant4 simulation from scratch and perform simple detector studies in time frames reasonable for graduation \cite{DurantiPriv}.

In the field of muon radiography \cite{Bonechi2020_1}, the MURAVES collaboration is performing geological introspections on 
volcanoes \cite{DErrico2020}. It uses a hodoscope made of scintillating bars as a muon tracker, to infer the amount of matter traversed by 
atmospheric muons by comparing the number of detected muons that cross the volcano against the number of muons coming from the free sky. GGS 
has been used for a first implementation of the full Monte Carlo simulations, making it possible for a very limited workforce (about 1-2 
persons) to quickly build up the full simulation stack. The standard GGS components have been efficiently integrated with custom components 
specifically developed for muon radiography \cite{Mori2014}, demonstrating the validity of the development approach described in section 
\ref{sect:requirements} despite GGS having later been replaced by PUMAS \cite{Niess2018} as the simulation framework of choice for the 
collaboration. However, other small muon radiography projects (\cite{Bonechi2020_2,Ambrosino2015}), have been able to profit from the GGS 
development work done for MURAVES and are still using GGS.

The CALET collaboration operates a detector for direct cosmic-ray detection onboard the International Space Station, with the main goal 
of providing accurate direct measurements of the spectra of electrons+positrons above TeV and of atomic nuclei up to hundreds of TeV
\cite{Maestro2020}. The main device is a deep, homogeneous electromagnetic calorimeter made of lead--tungstate bars, complemented by a 
preshower tracking device and a charge detector. The collaboration makes use of three different Monte Carlo codes based on EPICS 
\cite{EPICS}, FLUKA and Geant4 respectively, to better assess the effects of the systematics related to the physics models and 
approximations. The usage of GGS have helped in mitigating the impact of the fragmentation of a limited workforce (of the order of 5 
individuals) into three different projects, allowing a single person to build up and maintain the whole Geant4 simulation.

The HERD detector is aimed at measuring the spectrum and composition of cosmic rays at the highest energies ever reached by direct 
detection experiments, and will be installed on the future Chinese Space Station \cite{Zhang2014}. In order to significantly improve the 
performance with respect to the current generation of space-based instruments it makes use of innovative design choices, like the 
sensitivity on 5 detector sides and the usage of a 3D homogeneous calorimeter. The HERD collaboration is currently optimizing the detector 
design by exploring several configurations of its subsystems. GGS is currently used for the official Geant4 simulation for HERD; one of the 
benefits deriving from its adoption is the possibility to easily change the detector layout (e.g. replacing the silicon tracker with 
the fiber tracker) without the need to adjust the Root output format or the readout routines. It also allowed to introduce the simulation of 
transition radiation alongside the physics processes included in the standard Geant4 physics lists. The HERD use case proved the usefulness 
of GGS in speeding up design optimization studies also at medium detector scales and its capability to meet non-trivial customization 
requirements.

\section{Conclusions}
\label{sect:conclusions}
The Generic Geant4 Simulation software package can provide a significant help in developing Monte Carlo simulations of particle detectors 
based on Geant4. Its default implementations of mandatory user classes can in many cases cover most of the user needs in the early 
development stages, thus speeding up the initial deployment of the simulation code, while its extensible architecture and plugin system 
allows for a deep customization of almost all the simulation components, when the need for higher accuracy and dedicated features become 
important in the advanced stages of the experiment. A set of utilities like readout routines for output files and applications for event 
visualization and geometry conversion allows the user to quickly implement data analysis programs, to visualize the output of the simulation 
and to interface the geometry to other tools.

GGS has already been used with profit in contexts ranging from bachelor theses to space experiments, proving the effectiveness and 
robustness of its design and implementation.

\section{Ackowledgements}
The author acknowledges the contribution of V. Formato who developed the {\it GGSLeonard} event display application and ported the code 
to macOS. The author also thanks L. Pacini for fruitful discussions and suggestions, and M. Duranti for reading the manuscript and 
providing some use cases.

\bibliography{bibfile}

\begin{thebibliography}{10}
\expandafter\ifx\csname url\endcsname\relax
  \def\url#1{\texttt{#1}}\fi
\expandafter\ifx\csname urlprefix\endcsname\relax\def\urlprefix{URL }\fi
\expandafter\ifx\csname href\endcsname\relax
  \def\href#1#2{#2} \def\path#1{#1}\fi

\bibitem{Allison2016}
J.~Allison, et~al., Recent developments in {Geant4}, Nucl. Instrum. Meth. A 835
  (2016) 186--225.
\newblock \href {https://doi.org/10.1016/j.nima.2016.06.125}
  {\path{doi:10.1016/j.nima.2016.06.125}}.

\bibitem{Battistoni2015}
G.~Battistoni, et~al., Overview of the {FLUKA} code, Ann. Nucl. En. 82 (2015)
  10--18.
\newblock \href {https://doi.org/10.1016/j.anucene.2014.11.007}
  {\path{doi:10.1016/j.anucene.2014.11.007}}.

\bibitem{Faddegon2020}
B.~Faddegon, et~al.,
  \href{http://www.sciencedirect.com/science/article/pii/S1120179720300715}{{The
  TOPAS tool for particle simulation, a Monte Carlo simulation tool for
  physics, biology and clinical research}}, Physica Medica 72 (2020) 114 --
  121.
\newblock \href {https://doi.org/https://doi.org/10.1016/j.ejmp.2020.03.019}
  {\path{doi:https://doi.org/10.1016/j.ejmp.2020.03.019}}.
\newline\urlprefix\url{http://www.sciencedirect.com/science/article/pii/S1120179720300715}

\bibitem{Botta2013}
F.~Botta, et~al., {Use of the FLUKA Monte Carlo code for 3D patient-specific
  dosimetry on PET-CT and SPECT-CT images}, Physics in medicine and biology 58
  (2013) 8099--8120.
\newblock \href {https://doi.org/10.1088/0031-9155/58/22/8099}
  {\path{doi:10.1088/0031-9155/58/22/8099}}.

\bibitem{Lattuada2017}
D.~Lattuada, et~al., {A fast and complete GEANT4 and ROOT Object-Oriented
  Toolkit: GROOT}, EPJ Web Conf. 165 (2017) 01034.
\newblock \href {https://doi.org/10.1051/epjconf/201716501034}
  {\path{doi:10.1051/epjconf/201716501034}}.

\bibitem{Santin2006}
G.~Santin, V.~Ivanchenko, H.~Evans, P.~Nieminen, E.~Daly, {GRAS: a
  general-purpose 3-D Modular Simulation tool for space environment effects
  analysis}, Nuclear Science, IEEE Transactions on 52 (2006) 2294 -- 2299.
\newblock \href {https://doi.org/10.1109/TNS.2005.860749}
  {\path{doi:10.1109/TNS.2005.860749}}.

\bibitem{Lee2007}
K.~Lee, et~al., {Space Applications of the FLUKA Monte‐Carlo Code: Lunar and
  Planetary Exploration}, AIP Conference Proceedings 884~(1) (2007) 243--248.
\newblock \href {https://doi.org/10.1063/1.2710587}
  {\path{doi:10.1063/1.2710587}}.

\bibitem{Arce2014}
P.~Arce, et~al., {GAMOS}: A framework to do {Geant4} simulations in different
  physics fields with an user-friendly interface, Nucl. Instrum. Meth. A 735
  (2014) 304--313.
\newblock \href {https://doi.org/10.1016/j.nima.2013.09.036}
  {\path{doi:10.1016/j.nima.2013.09.036}}.

\bibitem{Aad2010}
G.~Aad, et~al., {The ATLAS Simulation Infrastructure}, Eur. Phys. J. C 70
  (2010) 823--874.
\newblock \href {https://doi.org/10.1140/epjc/s10052-010-1429-9}
  {\path{doi:10.1140/epjc/s10052-010-1429-9}}.

\bibitem{Clemencic2011}
M.~Clemencic, et~al., {The LHCb Simulation Application, Gauss: design,
  evolution and experience}, J. Phys. Conf. Ser. 331 (2011) 032023.
\newblock \href {https://doi.org/10.1088/1742-6596/331/3/032023}
  {\path{doi:10.1088/1742-6596/331/3/032023}}.

\bibitem{Siddi2019}
B.~G. Siddi, D.~Müller, {(LHCb collaboration)}, {Gaussino - a Gaudi-Based Core
  Simulation Framework}, in: {2019 IEEE Nuclear Science Symposium (NSS) and
  Medical Imaging Conference (MIC)}, 2019, pp. 1--4.
\newblock \href {https://doi.org/10.1109/NSS/MIC42101.2019.9060074}
  {\path{doi:10.1109/NSS/MIC42101.2019.9060074}}.

\bibitem{Lei2002}
F.~Lei, et~al., {MULASSIS: A Geant4-based multilayered shielding simulation
  tool}, IEEE Trans. Nucl. Sc. 49~(6) (2002) 2788--2793.
\newblock \href {https://doi.org/10.1109/TNS.2002.805351}
  {\path{doi:10.1109/TNS.2002.805351}}.

\bibitem{Santin2005}
G.~Santin, et~al., {GRAS: a general-purpose 3-D Modular Simulation tool for
  space environment effects analysis}, IEEE Trans. Nucl. Sc. 52~(10) (2005)
  2294--2299.
\newblock \href {https://doi.org/10.1109/TNS.2005.860749}
  {\path{doi:10.1109/TNS.2005.860749}}.

\bibitem{Zoglauer2006}
A.~Zoglauer, et~al., {MEGAlib: Medium Energy Gamma-ray Astronomy library}, New
  Astr. Rev. 50~(7-8) (2006) 629--632.
\newblock \href {https://doi.org/10.1016/j.newar.2006.06.049}
  {\path{doi:10.1016/j.newar.2006.06.049}}.

\bibitem{Antcheva2009}
I.~Antcheva, et~al., {ROOT — A C++ framework for petabyte data storage,
  statistical analysis and visualization}, Computer Physics Communications
  180~(12) (2009) 2499 -- 2512, 40 YEARS OF CPC: A celebratory issue focused on
  quality software for high performance, grid and novel computing
  architectures.
\newblock \href {https://doi.org/https://doi.org/10.1016/j.cpc.2009.08.005}
  {\path{doi:https://doi.org/10.1016/j.cpc.2009.08.005}}.

\bibitem{GDMLweb}
{Geometry Description Markup Language (GDML)}, \url{http://gdml.web.cern.ch}.

\bibitem{Petric2017}
M.~Petri\u{c}, et~al., {Detector simulations with DD4hep}, J. Phys. Conf. Ser.
  898~(4) (2017) 042015.
\newblock \href {https://doi.org/10.1088/1742-6596/898/4/042015}
  {\path{doi:10.1088/1742-6596/898/4/042015}}.

\bibitem{g4pub}
{Geant4 publications page}, \url{https://geant4.web.cern.ch/publications}.

\bibitem{g4val}
{Geant4 validation portal}, \url{https://geant-val.cern.ch/}.

\bibitem{Hrivnacova2008}
I.~Hrivnacova, B.~Viren, {The virtual geometry model}, J. Phys. Conf. Ser. 119
  (2008) 042016.
\newblock \href {https://doi.org/10.1088/1742-6596/119/4/042016}
  {\path{doi:10.1088/1742-6596/119/4/042016}}.

\bibitem{Tadel2010}
M.~Tadel, {Overview of EVE: The event visualization environment of ROOT}, J.
  Phys. Conf. Ser. 219 (2010) 042055.
\newblock \href {https://doi.org/10.1088/1742-6596/219/4/042055}
  {\path{doi:10.1088/1742-6596/219/4/042055}}.

\bibitem{GGSrepo}
{GGS git repository }, \url{https://baltig.infn.it/mori/GGSSoftware}.

\bibitem{Duranti2021}
M.~Duranti, et~al., \href{https://tinyurl.com/MDPITiming}{Advantages and
  requirements in time resolving tracking for astroparticle experiments in
  space}, solicited paper in preparation for a Special Issue “Timing
  Detectors” of Instruments (2021).
\newline\urlprefix\url{https://tinyurl.com/MDPITiming}

\bibitem{DurantiPriv}
M.~Duranti, {private communication}.

\bibitem{Bonechi2020_1}
L.~Bonechi, R.~D’Alessandro, A.~Giammanco, Atmospheric muons as an imaging
  tool, Reviews in Physics 5 (2020) 100038.
\newblock \href {https://doi.org/10.1016/j.revip.2020.100038}
  {\path{doi:10.1016/j.revip.2020.100038}}.

\bibitem{DErrico2020}
M.~D'Errico, et~al., {Muon radiography applied to volcanoes imaging: the
  MURAVES experiment at Mt. Vesuvius}, JINST 15~(03) (2020) C03014.
\newblock \href {https://doi.org/10.1088/1748-0221/15/03/C03014}
  {\path{doi:10.1088/1748-0221/15/03/C03014}}.

\bibitem{Mori2014}
N.~Mori, , et~al., {A Geant4 framework for generic simulations of atmospheric
  muon detection experiments}, Vol.~60, 2014.
\newblock \href {https://doi.org/10.4401/ag-7383} {\path{doi:10.4401/ag-7383}}.

\bibitem{Niess2018}
V.~Niess, A.~Barnoud, C.~C\^arloganu, E.~Le~M\'en\'edeu, {Backward Monte-Carlo
  applied to muon transport}, Comput. Phys. Commun. 229 (2018) 54--67.
\newblock \href {http://arxiv.org/abs/1705.05636} {\path{arXiv:1705.05636}},
  \href {https://doi.org/10.1016/j.cpc.2018.04.001}
  {\path{doi:10.1016/j.cpc.2018.04.001}}.

\bibitem{Bonechi2020_2}
L.~Bonechi, et~al., {Multidisciplinary applications of muon radiography using
  the MIMA detector}, JINST 15~(05) (2020) C05030.
\newblock \href {https://doi.org/10.1088/1748-0221/15/05/C05030}
  {\path{doi:10.1088/1748-0221/15/05/C05030}}.

\bibitem{Ambrosino2015}
F.~Ambrosino, et~al., {Assessing the Feasibility of Interrogating Nuclear Waste
  Storage Silos using Cosmic-ray Muons}, JINST 10~(06) (2015) T06005.
\newblock \href {http://arxiv.org/abs/1411.2382} {\path{arXiv:1411.2382}},
  \href {https://doi.org/10.1088/1748-0221/10/06/T06005}
  {\path{doi:10.1088/1748-0221/10/06/T06005}}.

\bibitem{Maestro2020}
P.~Maestro, et~al., {CALET Results after Three Years on Orbit on the
  International Space Station}, Phys. Atom. Nucl. 82~(6) (2020) 766--772.
\newblock \href {https://doi.org/10.1134/S1063778819660384}
  {\path{doi:10.1134/S1063778819660384}}.

\bibitem{EPICS}
{EPICS home page }, \url{https://cosmos.n.kanagawa-u.ac.jp/EPICSHome/}.

\bibitem{Zhang2014}
S.~N. Zhang, et~al., {The high energy cosmic-radiation detection (HERD)
  facility onboard China's Space Station}, Proc. SPIE Int. Soc. Opt. Eng. 9144
  (2014) 91440X.
\newblock \href {http://arxiv.org/abs/1407.4866} {\path{arXiv:1407.4866}},
  \href {https://doi.org/10.1117/12.2055280} {\path{doi:10.1117/12.2055280}}.

\end{thebibliography}

\end{document}